\documentclass[amsmath,amsfonts,aps,prb,preprint,showpacs]{revtex4}

\usepackage{graphicx}
\usepackage{dcolumn}
\usepackage{bm}
\begin{document}

\title{ Coherently photo-induced ferromagnetism in diluted magnetic
semiconductors} 

\author {J. Fern\'andez-Rossier, $^{1,2\ast}$ C. Piermarocchi
$^{3,4}$, P. Chen$^{3,5}$, A. H. MacDonald$^{1}$  and L. J. Sham$^{3}$}

\affiliation{
$^{1}$ Department of Physics, University of Texas at Austin. University Station
C1600, Austin, TX 78712 
\\$^{2}$ Departamento de F{\'\i }sica Aplicada,
Universidad de Alicante, San Vicente del Raspeig 03690, Alicante, Spain
\\$^{3}$  Department of Physics, University of California San Diego.9500 Gilman
Drive, La Jolla, CA92093   
\\$^{4}$ Department of Physics and Astronomy
Michigan State University. 4263 Biomedical and Physical Sciences East Lansing,
MI 48824-2320 
\\$^{5}$ Department of Chemistry University of California 
Berkeley.  406 Latimer Hall, Berkeley, CA 94720-1460  }

\date{\today}

\begin{abstract}
Ferromagnetism is predicted in undoped diluted magnetic semiconductors 
illuminated by intense {\it sub-bandgap} laser radiation . The mechanism for
photo-induced ferromagnetism is coherence between conduction and valence
bands induced by the light which leads to an optical exchange interaction. The
ferromagnetic critical temperature T$_C$ depends both on the properties of the
material and on the frequency and intensity of the laser and  could be
above $ 1 {\rm K}$.
\end{abstract}

\maketitle

Information processing in electronic devices is based on control of charge flow
in semiconductor materials, whereas non-volatile information storage exploits
ferromagnetism, spontaneous alignment of the spins of many electrons. {\em
Spintronics} \cite{Spintronics} aims to achieve a merger of these technologies,
motivating interest in new ferromagnetic semiconductors like (Ga,Mn)As and
other (III,Mn)V compounds \cite{Ohno}, and 
in diluted magnetic semiconductors (DMS) like
(Cd,Mn)Te and other (II,Mn)VI materials \cite{NatureMollenkamp} with free moments that can be
aligned by external fields. The optical properties of these semiconductors are
particularly interesting in this respect, because laser radiation can control
 both charge and spin dynamics\cite{opticS}. The possibility of optically controlling
semiconductors with dilute magnetic elements is now being explored actively
\cite{opticDMS}.

In this paper we predict a  qualitatively new effect, photo-induced
ferromagnetism, in which magnetic order is induced in an otherwise paramagnetic
(II,Mn)VI DMS by laser light that is  {\it below} the absorption edge. The
influence of the laser field on the electronic system is reactive,  rather than
dissipative and follows ultimately from a change in the  effective electronic
Hamiltonian analogous to those that yield dipole forces in atomic physics. The
II-VI parent compounds are intrinsic semiconductors with states of $p$
character at the top of the valence band and of $s$ character at the bottom of
the conduction band \cite{Furdyna}. They have a sizable interband optical matrix
element, $\vec{d}_{cv}=\langle v|e\vec{r}|c\rangle$. In (II,Mn)VI alloys,
Mn substitutes for the group-II atoms. The electronic structure of the external
$s,p$ shells of Mn and group II atoms is very similar, so that the conduction
and valence band are barely affected by moderate Mn doping. However, the $d$
shell of each Mn atom is only half full, in contrast to the full shells of the
atoms for which they substitute,  and forms a local moment with spin
$S=\frac{5}{2}$. These magnetic moments are the only low energy electronic
degrees of freedom in ideal (II,Mn)VI materials.

The magnetic moment of a Mn $d$ shell interacts with the spins of electrons in
both conduction and valence bands through the exchange coupling
\begin{equation}
 {\cal H}_{\rm exch}= \sum_{i,b} J_{b}\vec{M}_i \cdot
\vec{S}_{b}(\vec{R}_i), \label{exch}
\end{equation}
 where $\vec{M}_i$ refers to the Mn atom located at $\vec{R}_i$,
the spin density of the $b=e$ conduction electrons and $b=h$ valence band holes
is
$\vec{S}_{\rm b}(\vec{R}_i)$, and $J_{b}$ are the exchange coupling constants
\cite{Furdyna}. When the Mn atoms are spin-polarized this interaction leads, in a mean
field and virtual crystal approximation (MFVCA) whose approximate validity is
well established \cite{Furdyna,Dietl},  to an effective magnetic field $g \mu_B \vec{B}^{\rm
eff}\equiv J_{e,h} c_{\rm Mn} \vec{\cal M}$ 
experienced by the spins of the carriers,
where $c_{\rm Mn}$ is the density of Mn atoms and $\vec{\cal M}$ is their
average magnetization. This leads to band spin-splittings as large as 100~meV,
evident in optical spectra and known as {\em giant Zeeman splittings}
\cite{Furdyna}. 

In doped systems  band electrons  mediate RKKY interactions between  Mn spins. 
On the other hand, virtual fluctuations in the Mn valence  lead to short range
superexchange antiferromagnetic interactions which couple only nearest 
neighbors.  In a sample with a fraction $x$ of Mn randomly located at the
cation sites,  $x_{\rm eff}\equiv x(1-x)^{12}$ is the fraction of sites
containing Mn atoms without a magnetic first neighbor. 
Consequently, samples with a low Mn  concentration ($x\simeq 0.01$) and no
band carriers are paramagnetic.

It has been shown \cite{ORKKY} that two localized spins in a semiconductor  can
interact via the virtual carriers created in the (otherwise empty) bands by a
laser of frequency  smaller than the band gap. The effective interaction is a 
ferromagnetic Heisenberg coupling which we shall call optical RKKY (ORKKY). It
is our contention that the ORKKY interaction can drive a diluted magnetic
semiconductor  into  a ferromagnetic phase at low temperatures. 
In this paper, we
develop a microscopic theory for this new type of ferromagnetism and
we explore its properties.

When the material is illuminated by a laser with electric field $E_{\alpha}$,
and frequency $\omega_L$ below the semiconductor band gap, $E_g$, a reactive
dipolar energy is stored in the semiconductor: 
\begin{equation}{\cal E}=-\sum_{\alpha\beta}
\chi'_{\alpha\beta}(\omega_L)E_{\alpha} E_{\beta} 
\label{e1}
\end{equation}
 where $\chi'_{\alpha\beta}(\omega_L)$ is the real part of the retarded optical
response function. 
In this case, as in many others where direct absorption
processes are negligible, the laser field acts like a new tunable thermodynamic
variable which changes the properties of the system.
$\chi'_{\alpha\beta}(\omega)$ depends on the inter-band transition energies
which, due to the exchange interaction (\ref{exch}), depend in turn on the
collective magnetization $\vec{\cal M}$. In an illuminated sample, ${\cal E}$
{\em does} depend on $\vec{\cal M}$. As we show below, ${\cal E}$ is minimized
when the Mn spins are fully polarized and the system is ferromagnetic.
Importantly, because $\hbar \omega_L<E_g$, the dissipative part
$\chi''_{\alpha\beta}(\omega_L)$ of the response function is zero, so that no
real electron-hole pairs are created, and heating is strongly suppressed. 

We now derive ${\cal E}$ starting from the microscopic Hamiltonian.  The
relevant degrees of freedom are the electrons in the conduction band, holes in
the valence band and the Mn spins.   The exchange interaction (\ref{exch}) is
included in the VCMFA, which results in a spin splitting of the bands when the
Mn spins are polarized \cite{Furdyna,Dietl}.  
For simplicity, we have neglected
the important spin-orbit interaction influence on the valence bands;
although these interactions do not change the qualitative physics, they
do introduce a characteristic dependence on the laser light polarization that
we do not discuss here. A classical laser
field couples the valence and conduction bands. We only consider here 
the case of  unpolarized laser light, where the two
optically active (OA) interband transitions are driven with equal strength. 
We use the selection
rule for the heavy holes at $k=0$ so that an electron with spin $\sigma$ is
promoted from the valence band to the conduction band.  The resultant
magnetization comes from the cooperative effects of the Mn spins through their
interactions with the carriers, unlike optical orientation effects that can be induced
by polarized light in these materials. 

The Hamiltonian reads $H=H_c + H_v + H_L + V_C $ with: 
\begin{eqnarray}
H_c&=&\sum_{\sigma,\vec{k}} 
\left[\epsilon^e_{\vec{k}} - \frac{\sigma}{2} J_{e} c_{Mn}|\vec{\cal
M}|\right]c_{{\vec{k}},\sigma}^{\dagger}c_{\vec{k},\sigma}\nonumber \\
H_v&=&\sum_{\sigma,{\vec{k}}} 
\left[\epsilon^h_{\vec{k}} - \frac{\sigma}{2} J_{h} c_{Mn}|\vec{\cal
M}|\right]d_{{\vec{k}},\sigma}^{\dagger}d_{{\vec{k}},\sigma}\nonumber \\
H_L&=&\frac{\Omega}{2}
\sum c^{\dagger}_{{\vec{k}},\sigma} 
e^{i \omega_L t} d^{\dagger}_{{\vec{k}},\sigma} +h.c. \nonumber \\
V_C&=& \frac{1}{2{\cal V}}\sum_{\vec{q}} V(\vec{q}) \rho(\vec{q}) \rho(-\vec{q})
\label{hamil}
\end{eqnarray}
where we have chosen the spin quantization axis along the collective
magnetization axis $\vec{\cal M}$. The conduction and valence band dispersions
are $\epsilon^e_{\vec{k}}$ and $\epsilon^h_{\vec{k}}$ respectively. 
The strength of the light matter
coupling is quantified by the Rabi energy, $\Omega = d_{cv} E_{0}$ where $E_0$
is the amplitude of the electric field of the laser. The last term accounts for
the Coulomb interaction in the dominant term approximation with 
$\rho(q)=\rho_e(q)-\rho_h(q)$, 
$\rho_e(q)=\sum_{{\vec{k}},\sigma} 
c^{\dagger}_{{\vec{k}},\sigma} c_{\vec{k}+\vec{q},\sigma}$ 
$\rho_h(q)=\sum_{{\vec{k}},\sigma} 
d^{\dagger}_{{\vec{k}},\sigma} d_{\vec{k}+\vec{q},\sigma}$, 
$V(q)= \frac{4\pi}{\epsilon q^2}$ and ${\cal V}$ stands 
for the volume of the sample.
 
The dynamics of the carrier density matrix under the influence of  Hamiltonian
(\ref{hamil}) in the spinless case with $J_{e,h}=0$ has been 
studied previously by a number of authors, who treat Coulomb interactions in the
Hartree Fock (HF) approximation \cite{SR,Comte,Zimmermann}.   Numerical
solution of the dynamical equations \cite{Zimmermann} show that
the density matrix reaches a steady state due to the inhomogeneous broadening
provided by the different $k$ states. Spontaneous emission and other phenomena
will also contribute to the damping of the density matrix dynamics in a time
scale $T_2$ which is in the picosecond range. The HF steady state  can be described
as the vacuum of a dressed Hamiltonian \cite{Comte} and  resembles very much
a BCS state for excitons, where $\Omega$ and the electron hole (eh) interaction
act as the pairing force and $\delta_0\equiv E_g-\omega_L$  plays the role of
the chemical potential. When  $\delta_0$ is positive and sufficiently large,
the average density of band carriers vanishes identically when the laser is
switched off. In that limit, the wave function of the ground state of the
dressed Hamiltonian can be mapped to a coherent state of non-interacting bosonic
excitons \cite{SR} with  binding energy $\epsilon_x$, Bohr radius $a_B$ and
ground state wave function $\psi_{1s}(r)$ corresponding to the electron hole
Coulomb  potential $V(q)$ above. A crossover to the incoherent region, in which
the density of  carriers  can be non zero when the laser is switched off occurs
when $\delta_0$ decreases towards smaller or even negative values.  

When the spin degree of freedom is included, the equations for the two
optically active channels  decouple in the HF approximation \cite{spinfac} for
an unpolarized  laser.  The two sets of equations for the carrier
density matrix of the two OA channels, $+$ and $-$, are identical to those of
the spinless case \cite{SR,Zimmermann,Comte} except for the detuning, which
acquires a spin dependent magnetic shift. It proves useful to define a detuning
renormalized by the excitonic effects, $\delta\equiv\delta_0-\epsilon_x$ and
a spin dependent detuning
\begin{equation}
\delta_{\pm}\left({\cal M}\right)=
 E_g -\omega_L -\epsilon_x \pm {\cal J}\left({\cal M}\right) \equiv \delta \pm  
{\cal J}\left({\cal M}\right) 
\end{equation}
with  ${\cal J}({\cal M})\equiv (J_h-J_e)c_{\rm Mn}\frac{\langle {\cal M}\rangle}{2}$.
As a result of the magnetization, the detuning in one channel increases whereas
in the other it decreases. For ${\cal M}=0$  the coherent region (no absorption)
occurs for $\delta>0 $ \cite{Zimmermann}. In order to keep the semiconductor transparent in the
ferromagnetic phase, we need $\delta>{\cal J}\left({\cal M}\right)$. 
For the DMS considered below, the magnetic shift is large enough
so that the system is in the non-interacting exciton situation for which
analytical results are  available \cite{SR}. For instance,
the density of virtual excitons in each channel reads:
\begin{equation}
n_{\pm} a_B^3=
\frac{1}{4\pi}\frac{\Omega^2}{(\delta_{ \pm})^2}
\label{densi}
\end{equation}
This result is valid
provided that $n_{\sigma} a_B^3<0.1$, which is the case for all the results
presented below. This
condition imposes both a lower bound for $\delta$, which must be  
larger than 
$\delta_{\rm min}=|{\cal J}({\cal M}=S)|=J_{max}$, and an upper bound for
$\Omega$, which must be smaller than  $\Omega_{\rm max}=1.12 \times
(\delta-\delta_{\rm min})$. The former is  the condition for  the material to
remain non-absorbing in the ferromagnetic phase, whereas the latter is the
limit for the validity of the independent exciton approximation. In this limit
the reactive dipolar energy stored in the semiconductor because of the
interaction with the laser reads: 
\begin{equation} {\cal E}({\cal M} ) = 
-\frac{{\cal V}\Omega^2|\psi_{1s}(0)|^2}{4}
\left[\frac{ 1} {\delta_ +}+ \frac{ 1} {\delta_- }
\right]
\label{U}
\end{equation}
which should be compared with (\ref{e1}).
Formally eq. (\ref{U}) is the ground state energy of the dressed Hamiltonian, in
the non-interacting exciton limit. 
To explore the collective properties of the Mn magnetic moments under the
influence of the laser, we assume that the magnetic degrees of freedom are in
thermal equilibrium and that their internal energy ${\cal E}$ is given by Eq.
(\ref{U}). The magnetization is determined by the competition between the
dipolar energy density 
${\cal G}_{\cal E}\equiv \frac{1}{{\cal V}}\;{\cal E}({\cal M})$
  and the entropy  density in 
  ${\cal G}_S=-c_{Mn}k_b T{\cal S}({\cal M})$, where ${\cal S}({\cal M})$ is the
  entropy per spin as a function of the magnetization \cite{jfr-sham}. 
When the paramagnetic to ferromagnetic phase transition is continuous,  
we can expand
$G\left({\cal M}\right)$ around ${\cal M}=0$ to obtain an analytical formula 
for the Curie Temperature:
\begin{equation} k_B T_C = \frac{S(S+1)}{3} (J_e-J_h)^2 c_{\rm Mn}
\frac{|\psi_{1s}(0)|^2 \Omega^2}{4\delta^3}.
 \label{tc}
\end{equation} 
We note that the same equation can be derived  from the mean field theory using
the effective Heisenberg interaction defined by the optical exchange coupling
derived in ref. \onlinecite{ORKKY}. From equation  (\ref{tc}) and  the
numerical evaluation of $T_C$ shown below, we conclude that the important
factors to achieve sizable  Curie Temperature are compact excitons (small
$a_B$) and large Rabi energy $\Omega$.  From this point of view, (Zn,Mn)S and
(Zn,Mn)Se seem to be the most promising in the II-VI family, due to their small
Bohr radii.

In figure 1 we plot  ${\cal G}_{\cal E}(\cal{M})$ and  ${\cal G}_S(\cal{M})$
for bulk Zn$_{0.988}$Mn$_{0.012}$S ($x_{\rm eff}=0.01$) under the influence of
a laser with Rabi energy $\Omega=$ 5 meV and a detuning $\delta=41$ meV
\cite{note2}. For this concentration, the magnetic shift for the fully polarized case
is $J_{\rm max}=$21 meV. The upper panel of figure 1-a shows ${\cal G}_S(\cal M)$,  which
favours magnetic disorder ($\cal M$=0). In the middle panel of figure 1-a we see
how the dipolar energy is minimized for a maximal magnetic order $M=5/2$. In
figure 1-b we plot the total Landau functional $G\left({\cal
M}\right)={\cal G}_{\cal E}+{\cal G}_S$, for a set of temperatures between 105
and 115 mK. For each temperature, the order parameter $\cal M$ is obtained by
minimization of $G\left({\cal M}\right)$ (magenta points in figure 1b). The
curve ${\cal M}(T)$ so generated is shown in the lower panel of figure 1-a. A
discontinuous phase transition from paramagnetic (${\cal M} =0$) to ferromagnetic
(${\cal M}\neq 0$) occurs at $T_C=114$ mK. In that panel we also show 
${\cal M}(T)$ for some other detunings, $\delta=26$ meV ($T_C=$778 mK), $\delta=71$ meV
($T_C=$22 mK). We see that $T_C$ is a decreasing function of $\delta$ and 
that the shape of the ${\cal M}(T/T_C)$ curve varies: our model predicts 
continuous and discontinuous phase transitions in different regions of the phase diagram.  
Conveniently, equation (\ref{tc}) remains a good
approximation for $T_C$ even when the transition is discontinuous.

The Curie temperature as a function of $\delta$ and $\Omega$ is shown in figure
2-A for Zn$_{0.988}$Mn$_{0.012}$S. From figure 2B it is apparent that
ferromagnetism can be switched on and off, at a given temperature, by adjusting
the intensity or the frequency of the laser.  For a laser field with Rabi
energy $\Omega=10$ meV and $\delta =30.5$ meV our theory predicts $T_C= 1.5$
Kelvin. Continuous wave laser excitation with  Rabi splitting $\Omega=0.2$ meV
has been achieved \cite{smallrabi} which would yield $T_c\simeq$ 50 mK for bulk
Zn$_{0.988}$Mn$_{0.012}$S and $\delta=21.6$ meV \cite{note3}. In the opposite
limit, values of $\Omega$ as large as 1.5 eV can be achieved with 5 femtosecond
laser pulses \cite{largerabi}.  However, in order to observe ferromagnetic
order the laser pulse must be longer than the transition switching time,  which
should presumably have the same order of magnitude  as  $T_1$, the Mn
longitudinal relaxation time. In the absence of laser driven interactions,
this quantity depends strongly on $x$, the Mn concentration,  indicating that
the superexchange interactions are the dominant relaxation mechanism \cite{T1}.
For $x\simeq 0.01$, the laser driven interaction is dominant and presumably as
efficient as the superexchange interactions in samples with larger Mn
concentrations, where $T_1$ can be of the order of $0.2$ ns
\cite{opticDMS,T1}.  In contrast, the ferromagnetic  {\em interactions} will
switch on and off as fast as the laser pulse, on a femtosecond time scale with
state of the art laser techniques.

The full phase diagram in  $T$ and $\delta$  would have six distinct regimes
(see figure 2C).  For $\delta>J_{\rm max}$ the semiconductor is always
transparent and the system can undergo the purely coherent ferromagnetic phase
transition at a $T_C$  close to the estimates plotted in figure 2A. For
0$<\delta<J_{\rm max}$, the system is transparent and paramagnetic at high
temperature, but has the potential to become  ferromagnetic and absorbing at
low temperatures. In this situation, the system might return to the
paramagnetic phase, because of heating, or might be ferromagnetic (due to the
standard RKKY interaction mediated by the real photocarriers). Finally, for
negative detuning the system is always absorbing, and photo-induced carriers
can  lead to ferromagnetism through the same mechanisms that apply for
equilibrium carriers, leading to a transition at 
$
k_B T_C = \frac{S(S+1)}{3}
 \frac{2 n^{1/3}c_{\rm Mn}}{\hbar^2 (3\pi^2)^{(2/3)}} \left(J_{sd}^2 m_c+
J_{pd}^2 m_v \right)
$
 where $n$ is the  density of photocarriers.
Ferromagnetism induced by (real)  photocarriers  has been observed in (Cd,Mn)Te
heterostructures  at 2 Kelvin \cite{Boukari}

In summary, we propose that a diluted magnetic semiconductor exposed to a laser
below the absorption threshold will undergo a ferromagnetic phase transition at
low temperatures. The ferromagnetism takes place because the reactive dipolar
energy induced in the semiconductor is minimized by the ferromagnetic
configuration. Microscopically, the spins interact with each other via an
optical RKKY interaction mediated by virtual carriers \cite{ORKKY}. The
ferromagnetic interactions are switched on and off as fast as the laser.  The
mechanism of this laser induced ferromagnetism relies on interband optical
coherence and is radically different from other types of carrier mediated
ferromagnetism.

We thank T. Dietl for fruitful discussions.  This work has been supported by
the Welch Foundation, the Office of  Naval Research under grant N000140010951,
DARPA/ONR N0014-99-1-1096,  NSF DMR 0099572, DMR-0312491  and Ram\'on y Cajal
program (MCYT).  This has been partly funded by FEDER funds.   Ministerio de
Ciencia y Tecnología, MAT2003-08109-C02-01.

\newpage

{\bf CAPTION Figure 1}

Energy contributions as a function of $\cal M$.  Figure
1-A. Upper panel: ${\cal G}_S({\cal M})$ for two different temperatures. Middle
panel: $\frac{{\cal E}({\cal M})}{V}$ for $\delta=41$ meV and $\Omega=5 $ meV.
Lower panel: ${\cal M}(T/T_c)$  for $\Omega=5 $ meV and three different
values of $\delta$. Figure 1-B: total Landau energy ${\cal G}({\cal M})$ for
the Mn, for a  set of temperatures nearby $T_c$ for $\delta=41$ meV and
$\Omega=5$ meV.
 
{\bf CAPTION Figure 2}

(A). Contour map for transition temperature as a
function of the Rabi energy $\Omega$ and the detuning $\delta$,  for
(Zn$_{0.99}$,Mn$_{0.01}$)S. White region: outside validity range of linear
response. (B) ${\cal M}(\Omega,\delta)$ at a fixed temperature, $k_B T$=0.5
Kelvin for 
Zn$_{0.988}$Mn$_{0.012}$S. (C)  Schematic ($k_B T$, $\delta/J_{\rm
max}$) phase diagram, for fixed laser intensity. In the left  region
 ($\delta<0$) (right region ($\delta>J_{\rm max}$) ) the system is always absorbing
(coherent).  The red line in the coherent side signals the
ferromagnetic instability described in this
work. The red line in the absorbing part corresponds  to photocarrier mediated
ferromagnetism  $T_c$ \cite{Dietl,Boukari}.  The middle region shows a possible transition from
a paramagnetic-coherent  towards a ferromagnetic-absorbing phase.

\newpage

\begin{figure}
\includegraphics[width=10cm]{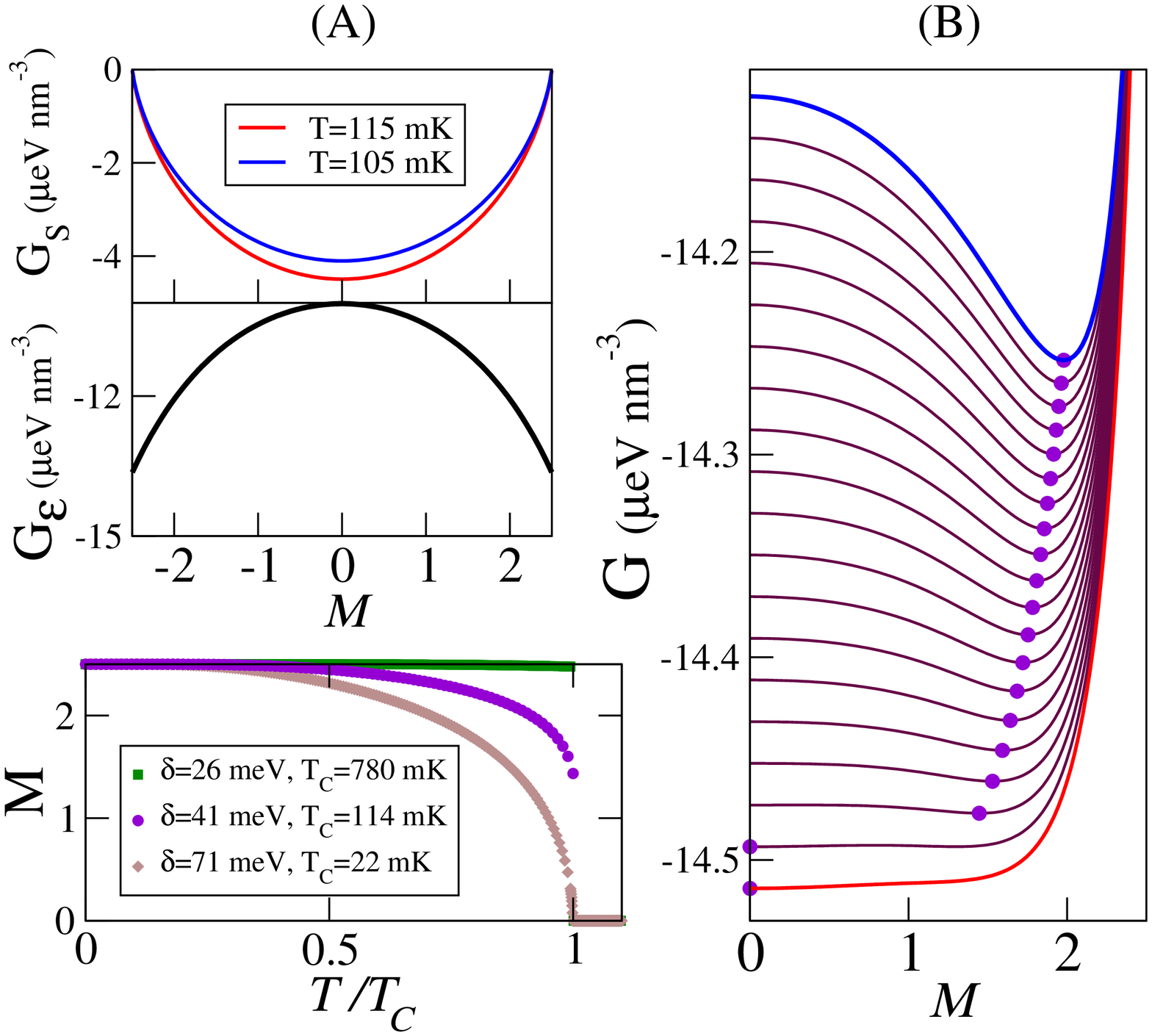} 
 \end{figure}

\newpage{Figure 1}
\pagebreak
\begin{figure}
\includegraphics[width=8cm]{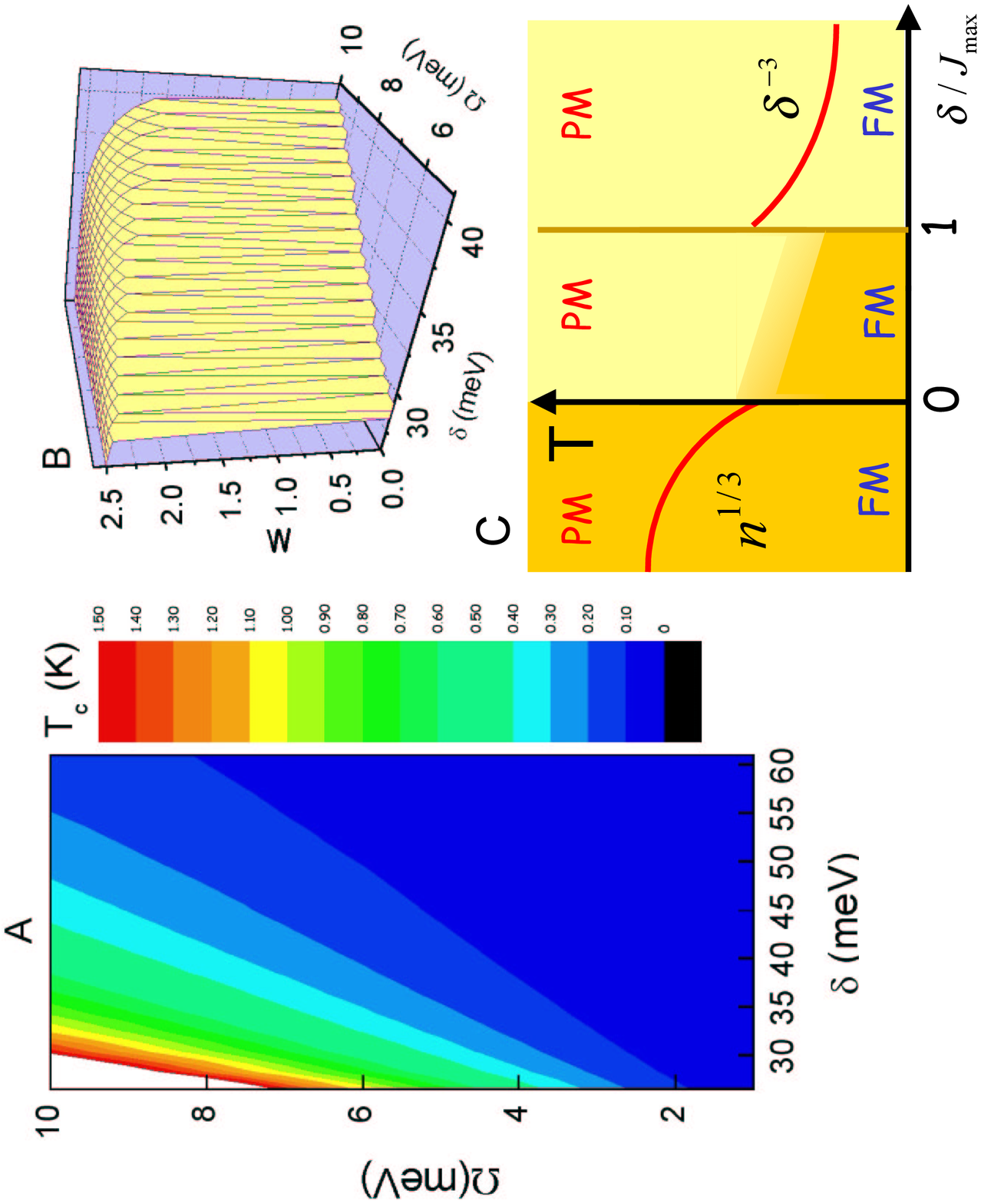} 
\end{figure}


\begin{references}
\bibitem{note2}
For Zn$_{0.988}$Mn$_{0.012}$S we take the exciton Bohr radius,
$a_B=$21 $\AA$ and the exchange constants $J_e=$ 7.8 eV $\AA^3$
($\alpha$=0.2 eV) and $J_h=$ -59.04 eV $\AA^3$ ($\beta=$ -1.5 eV).
\end{references}
\end{document}